# Probing the Nature of Defects in Graphene by Raman Spectroscopy

Axel Eckmann[1], Alexandre Felten[2,3], Artem Mishchenko[4], Liam Britnell[4], Ralph Krupke[3], Kostya S. Novoselov[4], Cinzia Casiraghi[1,2]

1 School of Chemistry and Photon Science Institute, University of Manchester, UK

2 Physics Department, Freie Universität Berlin, Germany

3 Karlsruhe Institute of Technology, Karlsruhe, Germany

4 School of Physics and Astronomy, University of Manchester, UK

Email: cinzia.casiraghi@manchester.ac.uk

**Abstract**

Raman Spectroscopy is able to probe disorder in graphene through defect-activated peaks. It is of great interest to link these features to the nature of disorder. Here we present a detailed analysis of the Raman spectra of graphene containing different type of defects. We found that the intensity ratio of the D and D' peak is maximum (~ 13) for $sp^3$-defects, it decreases for vacancy-like defects (~ 7) and reaches a minimum for boundaries in graphite (~3.5).

**Manuscript text**

Since its first experimental observation, graphene has triggered enthusiasm in the world's scientific community due to its outstanding properties.[1,2] In particular, near-ballistic transport at room temperature and high carrier mobilities[3-6] make it a potentially attractive material for nano-electronics.

Despite being praised for being inert, ultra-strong and impermeable to any gaseous material,[7] realistic graphene structures always contain defects.[8] One generally refers to defects in graphene as anything that breaks the symmetry of the infinite carbon honeycomb lattice. Thus, different types of defects can be defined such as edges, grain boundaries, vacancies, implanted atoms and defects associated to a change of carbon-hybridization, for example from $sp^2$ into $sp^3$. The amount and nature of defects strongly depend on the production method and may change from sample to sample. Both the amount and the nature of defects can have a strong influence on the properties of graphene samples[9] and can strongly vary with the graphene production and processing methods. For example, resonant scatterers, e.g.

atomic-sized defects that introduce mid-gap states close to the Dirac point, have been identified as the major limitation of electron mobility for graphene deposited on substrates.[10,11] On the other side, the control of the location of defects and their arrangement into ordered and extended structures allows making preparation of new graphene-based materials with novel properties.[12] Extended line defects could be used to guide charge as well as spin, atoms and molecules.[13] Defects also have strong influence on the chemical reactivity.[14] This makes defected graphene a prospective catalyst.[14] It is therefore of fundamental importance to be able to probe defects and to establish the precise nature of disorder.

Raman Spectroscopy is a well-established technique for investigating the properties of graphene.[15,16] This technique is able to identify graphene from graphite and few-layers graphene and it is sensitive to defects, excess charge (doping), strain and to the atomic arrangement of the edges.[15-24] Raman spectroscopy is able to probe defects in graphitic materials because, along with the G and 2D (also called G', being symmetry allowed) peaks that always satisfy the Raman selection rule, the Raman- forbidden D and D' bands appear in the spectrum.[25] They are activated by a single-phonon inter-valley and intra-valley scattering process, respectively, where the defect provides the missing momentum in order to satisfy momentum conservation in the Raman scattering process.[26-29]

Graphene is an ideal material to study defects because its 2D-nature makes it easy to add, remove or move carbon atoms, i.e. to introduce only a specific type of defect, in contrast to graphite or carbon nanotubes. Graphene is then the perfect target to investigate the sensitivity of the Raman spectrum on the nature of defects and finally build up a complete theory linking the Raman peak intensities to the number and type of defects.

The study of the evolution of the intensities of the Raman peaks for increasing disorder has been recently reported for vacancies-type defects,[30-32] but no analogue experimental work has been done on graphene with other types of defects. Here we study the Raman spectrum of a large amount of defected graphene samples, where different type and amount of defects have been introduced. $sp^3$-defects were introduced by fluorination and mild oxidation and compared to vacancy-like defects produced by $Ar^+$-bombardment.[30-32] Pristine defected graphene produced by anodic bonding[33] were also analyzed. Note that little is known about defects produced by this method.

We will show that the evolution of the Raman spectrum for increasing disorder depends on the type of defect and this is reflected in the defect-activated Raman intensities. In particular, our results clearly show that the intensity ratio between the D and D' peak is able to probe the nature of the defects for moderate amount of disorder. We then applied our new finding to defected graphene produced by anodic bonding: we found that these samples mainly contain vacancy-like defects. Thus, defected graphene produced by anodic bonding is very different from graphene chemical derivatives obtained by partial fluorination or oxidation. This has been further confirmed by using Atomic Force Microscopy (AFM) in tapping and conductive mode.

Chemical derivatives obtained by partial oxidization, hydrogenation and fluorination of graphene were used to investigate $sp^3$-type defects. Graphene samples were prepared by micro-mechanical exfoliation of single-crystal graphite flakes[34] (Nacional de Graphite LTDA) on $Si/SiO_x$. Graphene flakes were placed in a purpose-built chamber where they undergo an inductively coupled plasma at RF of 13.56 MHz.[35] The plasma treatments were performed at a power of 10 W and a pressure of 0.1 Torr. The amount of defects was tuned by changing the treatment time (between 2s and 300s). A controlled flow of oxygen, dihydrogen and tetrafluoromethane inside the chamber enabled graphene partial oxidization, hydrogenation or fluorination, respectively. In addition, partial fluorination was realized on separate exfoliated flakes, using the technique described in Ref. 36. Exfoliated flakes with no initial D peak were used.

Anodic bonding onto glass substrate[33] was used to prepare pristine graphene with defects. The quality of such flakes depends on the deposition parameters- we purposely choose deposition parameters that yield flakes with high D peak.[33]

Micro Raman measurements were performed with a confocal Witec spectrometer equipped with a 514.5nm (2.41eV) laser in backscattering configuration. We used a 100x objective giving a laser spot size of about 400nm. The laser power was kept well below 1mW to avoid damage or heating, which could induce desorption of the adatoms from graphene. The spectral resolution is ~ 3 $cm^{-1}$. The spectrometer is equipped with a piezoelectric stage that allows Raman mapping of area up to 200 x 200 $\mu m^2$. Because of the inhomogeneity of the fluorinated and anodic-bonded flakes, we used Raman mapping to collect a large amount of spectra with different amount of disorder. The D, G and 2D peaks are fitted with Lorentzian functions and the D' peak by a Fano lineshape. Here, we refer to peak intensity as the height of the peaks and it will be denoted as I(D), I(G), I(D'), I(2D) for the D, G, D' and 2D peaks, respectively.

We used AFM to further investigate the nature and morphology of defects. Tapping mode AFM was used to study anodic bonded samples, while conductive AFM was utilized to gather information on defects in fluorinated and oxidized graphene. Topography and current images were obtained with Atomic Force Microscope Nanoscope Dimension V (Bruker) in contact mode with conductive Pt/Ir coated cantilevers PPP-CONTPt (Nanosensors). This technique provides information on local conductivity, and can be used to distinguish the patches of $sp^3$ carbon (typically insulating) within a perfect graphene matrix.[37] Current was measured at fixed bias of 0.1 V applied to the tip via Keithley2400 source-meter. The fluorinated graphene substrate was grounded through 1 MΩ limiting resistor and the voltage drop across this resistor measured by 2184A Nanovoltmeter (Keithey) was measured as a function of tip position. Images were obtained in ambient conditions with scan rate 0.2 Hz and applied force of about 5 nN.

Figure 1(a) shows the Raman intensities measured on an oxidized graphene flake for increasing plasma treatment time. At short exposure time, i.e. for small defect concentrations, I(G) is practically constant, while I(D) and I(D') strongly increase with exposure time. At a certain defect concentration I(D) reaches a maximum and then starts decreasing. On the other side I(D') stays constant. Note that at high defect concentration the D' peak starts to merge with the G peak, so it is difficult to separate the individual contribution of the G and D' peak. Our observations agree with the results reported for graphene and multi-layer graphene bombarded with $Ar^+$ ions.[30-32] By using the same terminology introduced in disordered carbons,[38] the authors of these works have shown that the ratio I(D)/I(G) follows a two-stage evolution. By introducing a typical length $L_d$, representing the mean distance between two defects, they observed the following: i) at low defect concentration, I(D)/I(G) ~ $1/L_d^2$ (Stage 1); ii) at high defect concentration, I(D)/I(G) ~ $L_d^2$ (Stage 2). The transition between Stage 1 and 2 is usually observed at I(D)/I(G) ~ 3 at 2.41 eV (corresponding to $L_d$ ~ 2-5 nm).[32] In our case, the transition is observed at about 60s, corresponding to I(D)/I(G) ~ 4 (at 2.41eV). What exactly happens to the Raman intensities for $L_d$ ~ 0 still remains an open question. Ref. 32 proposes I(D)/I(G) ~ 0.8 for $L_d$ ~ 0, while Ref. 31 claims that I(D)/I(G) ~ 0 for $L_d$ ~ 0. They both analyzed samples with vacancies. In the case of $sp^3$-defected graphene, for example obtained by fluorination, the intensity of the D peak is never null, even in the most fluorinated samples.[36] In this case both the D and G peak intensities strongly decrease, so that the minimum I(D)/I(G) measured in stage 2 is ~ 0.8. This disagrees with the theory presented in Refs. 31,32, where the authors claim that this number should be dependent on the geometry of the defect. On the other side, in the highly disordered regime, the defects should be so close to each other and so many that the information about the geometry of the single defect should be lost. Thus, this regime needs further investigation.

Note that in some cases the integrated area (A) is used as intensity. It is then interesting to compare the evolution of I and A for increasing disorder. Figure 1 (b) shows that in Stage 1 both the Raman fit parameters follow the same evolution. Thus, in this range, the use of integrated area or amplitude is equivalent. This has been already observed for the G and the D peaks in ion bombarded graphene.[39] A difference is observed only in stage 2 because the decrease in intensity is compensated by an increase in the FWHM.[39] Here we show that this observation is also valid in the case of $sp^3$ type defects. Since our results mainly concern the low defect concentration range, in the following we will always refer to intensity as amplitude.

Figure 1b shows that there is a fundamental difference between the integrated intensity of defect-activated D and D' peaks and the two-phonon Raman line (2D). The theory predicts the double-resonant peaks to be strongly sensitive to the dynamics of the photo-excited electron-hole pair,[28] that is, to the scattering process it can undergo. In particular, any increase of defects will affect the electron lifetime, which translates in a decrease of the intensity. This is valid for D, D' and 2D peaks. However, in the case of the D and D' peaks, there is a further dependence: the D and D' intensities are also directly proportional to the defects concentration $n_d$.[29] This gives rise to the different behavior of the Raman peak intensities: in Stage 1, the D and D' peak intensity increase with increasing amounts of defects, while the 2D intensity stays almost constant. Thus, we expect D and D' to be proportional to each other. In stage 2, the effect of the reduced electron life-time dominates, so the integrated intensity of the peaks changes compared to Stage 1. Note that the D' peak integrated intensity does not decrease as seen for the D peak, in contrast to what expected by the resonant theory. This leads to a more complicated relation between I(D) and I(D'), which are no longer proportional to each other as in Stage 1. On a microscopic picture, the intensity of the defect-activated peaks starts decreasing when the average length an electron/hole travels in between two scattering events with a defect becomes smaller than the average length an electron/hole couple travels before scattering with an optical phonon.[30,31]

Figure 2 shows two representative Raman spectra of a fluorinated flake and pristine defected graphene, obtained by anodic bonding. The difference between the two spectra is visible at first sight: both spectra show well visible D and D' peaks and their combination mode (D+D' peak). In both cases I(D)/I(G) is ~ 2.3 and the G peak FWHM ~ 24 cm$^{-1}$, i.e. the two defected graphene samples belong to Stage 1. The D, G, 2D and D+D' peaks intensity almost perfectly match between the two spectra. In contrast, the D' peak intensity is very different: fluorinated graphene has a higher I(D)/I(D'), compared to pristine defected graphene. Figure 2 suggests that I(D)/I(D') could be used to identify the nature of defects.

We then performed a systematic analysis of the Raman spectra of all our samples, by including also some data from literature: oxidized graphene from Ref. 40, ion-bombarded graphene from Ref. 30-32 and poly-crystalline graphite from Ref. 41.

Figure 3 plots I(D)/I(G) versus I(D')/I(G) for all the samples. If we follow a disordering trajectory, i.e. we move from Stage 1 to Stage 2, we can observe that in Stage 1: I(D)/I(G) and I(D')/I(G) always increase. Taking into account that in this stage I(G) is constant, I(D') is simply proportional to I(D), inset Figure 3, as expected. However, Figure 3 shows that the proportionality factor depends on the type of samples: all the sp$^3$-type defected graphene (partially hydrogenated, fluorinated and oxidized graphene) share the same slope in the plot I(D)/I(G) vs I(D')/I(G), i.e. they have the same intensity ratio I(D)/I(D'). In contrast, defected graphene samples produced by ion-bombardment show a smaller I(D)/I(D'). Finally, poly-crystalline graphite, where the defect is given by the grain boundaries, shows an even smaller I(D)/I(D') (inset Fig. 3).

Our results can be easily explained by following the resonant Raman theory: in Stage 1, I(D) ~ A$_d$ n$_d$ and I(D') ~ B$_d$ n$_d$, where A$_d$ and B$_d$ are two constants, which both depend on the type of perturbation introduced by the defect in the crystal lattice, i.e. they depend on the nature of the defect.[29] Consequently, I(D)/I(D') ~ A$_d$/B$_d$, i.e. this parameter should not depend on the defect concentration, but only on the type of defect.

By fitting the data in Fig. 3, we found that I(D)/I(D') is maximum (~13) for defects associated with sp$^3$ hybridization, it decreases for vacancy-like defects (~7) and reaches a minimum for boundary-like defects in graphite (~3.5). This shows that I(D)/I(D') can be used to get information on the nature of defects. This makes Raman spectroscopy a powerful tool to fully characterize disorder in graphene.

Note that only a few works paid attention to the D' peak.[30,42] In general this peak is not much studied because of its relatively small intensity compared to the D peak, i.e. often the peak appears just as a small shoulder of the G peak. However, at moderate defect concentration, the D' peak can be clearly distinguished from the G peak and it can have relatively large intensity (up to 1/3 of the intensity of the G peak). A few examples are provided in the Supporting Information.

It is now interesting to compare our results with recent ab-initio calculations which simulate graphene with specific type of defects.[29] In particular, three idealized defects have been simulated: i) hopping defects, produced by the deformation of the carbon-bond; ii) on-site defects, which describe out-of-plane atoms bonded to carbon atoms (namely $sp^3$ hybridization); iii) charged impurities, describing any charged atom or molecule adsorbed over the graphene sheet. These defects are not expected to give detectable D and D' peaks.[29] The calculations show that in Stage 1 hopping defects should have $I(D)/I(D') \sim 10.5$, while on-site defects should be characterized by $I(D)/I(D') \sim 1.3$.[29] Hopping defects should describe defected graphene containing vacancies, while on-site defects should describe $sp^3$-defected graphene. However, experimentally, we found not only different numbers, but also that $I(D)/I(D')$ should be larger for hopping defects than on-site defects. The discrepancy between theory and experiments can be attributed to the idealized description of defects in the ab-initio calculations. It is well known, for example, that a real $sp^3$-defect cannot be described as a on-site defect only. This defect is expected to have both on-site and hopping components since the out-of-plane bonding with the atom also introduces distortions in the crystal lattice.[36,43] Furthermore, this type of defect is usually not isolated (as assumed in the ab-initio calculations), but it appears in form of dimers or clusters.[44]

In Stage 2 the Raman fit parameters do not show a clear dependence on the type of defect. This is probably because the defect concentration is so high that any information about the nature of the defect is lost. Note that the exact transition between Stage 1 and 2 seems to slightly change with the type of defect: the higher $I(D)/I(D')$, the higher is $I(D)/I(G)$ at which the transition from stage 1 to stage 2 is observed. This agrees with the theory of Ref.29, where "less damaging" defects should have higher "critical" defect concentration, at which the D line intensity starts to decrease. This could explain why Stage 2 cannot be achieved by grain boundaries in graphite.[30,31]

Interestingly, Figure 3 also shows that pristine defected graphene produced by anodic bonding have the same $I(D)/I(D')$ of ion-bombarded graphene. This should indicate the presence of vacancy-like defects in these samples. Thus, we do expect to see different defects in partially fluorinated graphene and defected graphene produced by anodic bonding. Therefore, we investigated topography and conductive AFM images of these two types of samples.

Figure 4(a) and (b) show two AFM topography images of defected graphene produced by anodic bonding. Figure 4(a) evidences the presence of holes distributed in a random pattern. The typical size of the holes is found to be ~80 nm, as can be seen in Figure 4(b).

Figure 4(c) and (d) are topography and conductive AFM images, respectively, taken on partially fluorinated graphene. The current measured between the conductive AFM tip and the contacted graphene sample deposited on an insulating substrate clearly indicates the presence of regions with lower conductivity (dark spots in Figure 4(d)). Several scans taken on the same area revealed that the pattern was reproducible (Figure S3, Supplementary Information). This observation rules out noise as the cause of conductivity variations. Comparison with the topography scan of the same area (Figure 4(c)) showed no strong correlation between height and conductivity. Thus, the low-conductivity "patches" can be attributed to fluorinated clusters. The typical size of the $sp^3$ clusters were ~20-30nm. The non-zero conductivity of the fluorinated clusters can be accounted for by their small size compared to the size of the tip and probably by the presence of tunnelling current. The AFM analysis shows that defected graphene produced by anodic bonding is characterized by the presence of holes, with sizes typically below 100nm, while fluorinated graphene exhibits $sp^3$ clusters of 20-30nm in size.

To conclude, this work offers a detailed analysis of the Raman peak intensities in defected graphene. By comparing samples with different amounts and type of defects, we have shown that I(D)/I(D') can be used experimentally to get information on the nature of defects in graphene. This makes Raman Spectroscopy a powerful tool to fully characterize disorder in graphene.

SUPPLEMENTARY INFORMATION: further measurements by conductive AFM and Raman spectra with corresponding fit lines of graphene containing different amount and type of defects. This material is available free of charge via the Internet at http://pubs.acs.org.

ACKNOWLEDGMENT
The authors thank F. Mauri for useful discussions. AF and CC acknowledge support from the Humboldt Foundation.

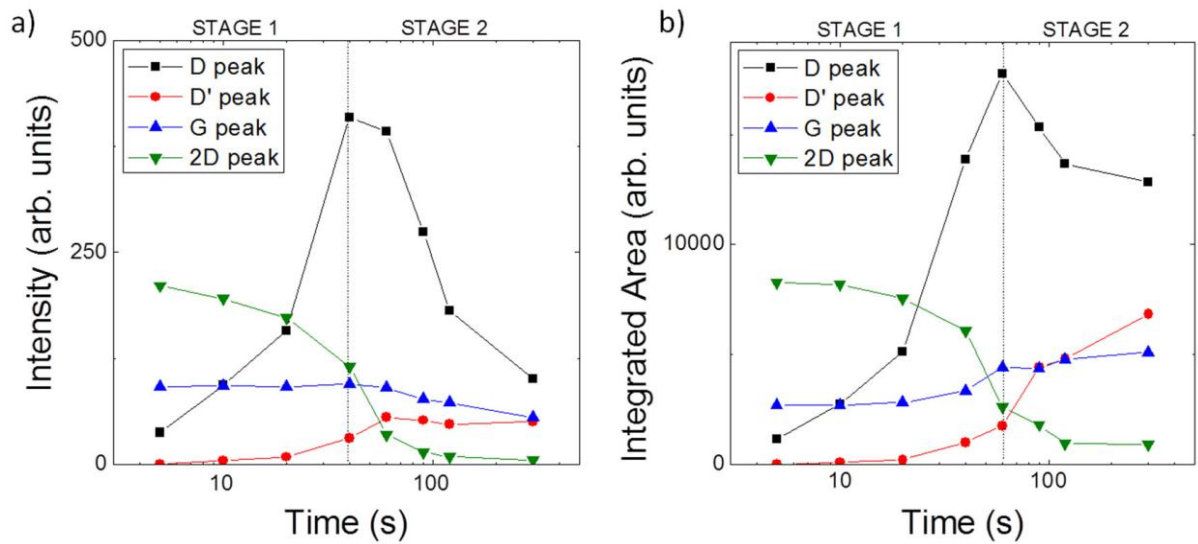

Figure 1 (color on-line) Raman intensity as amplitude (a) and integrated area (b) of oxidised graphene under increasing plasma exposure. Note that the units are arbitrary, i.e. we can compare the trend of the different peaks with exposure time, but we cannot compare the absolute numbers for a fixed time. The absolute intensity strongly depends on the experimental setup, such as the spectrometer sensitivity. This dependence is not included in these plots.

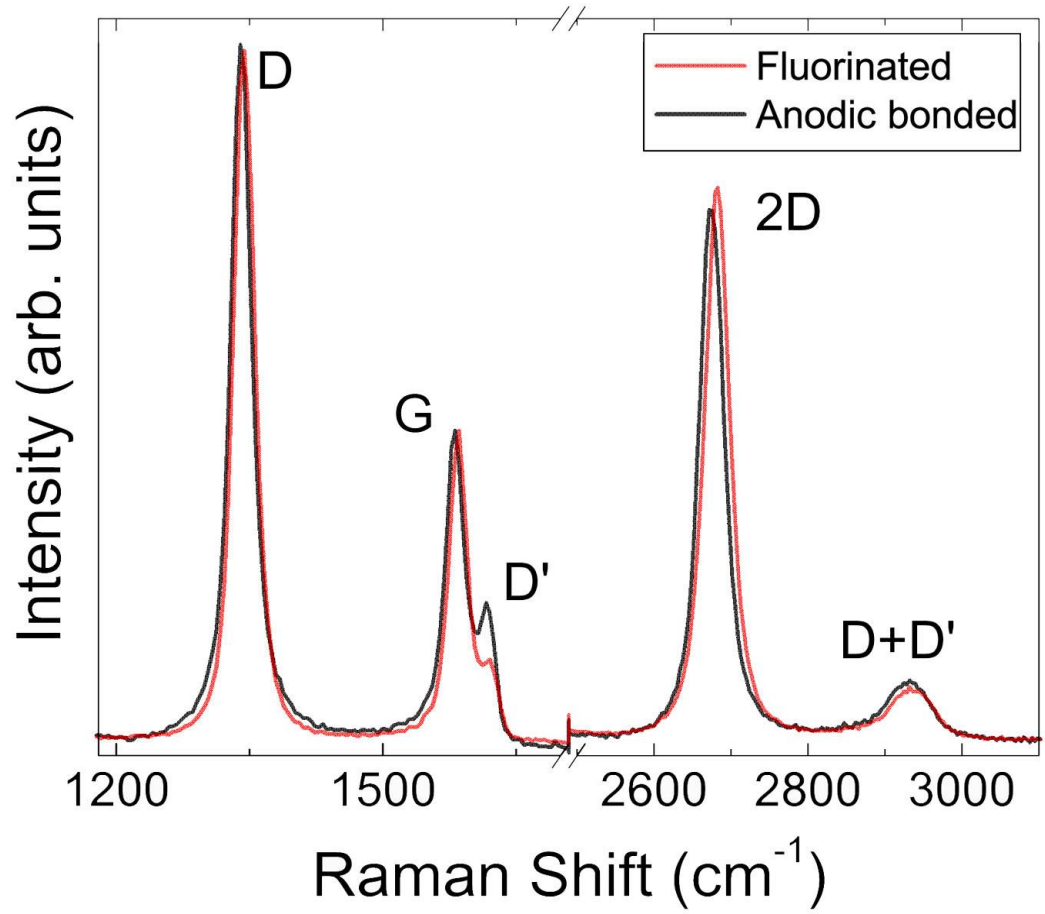

Figure 2 (color on-line) Raman spectrum of fluorinated (red) and defected graphene produced by anodic bonding (black), showing the same D, G and 2D intensities, but different D' intensity.

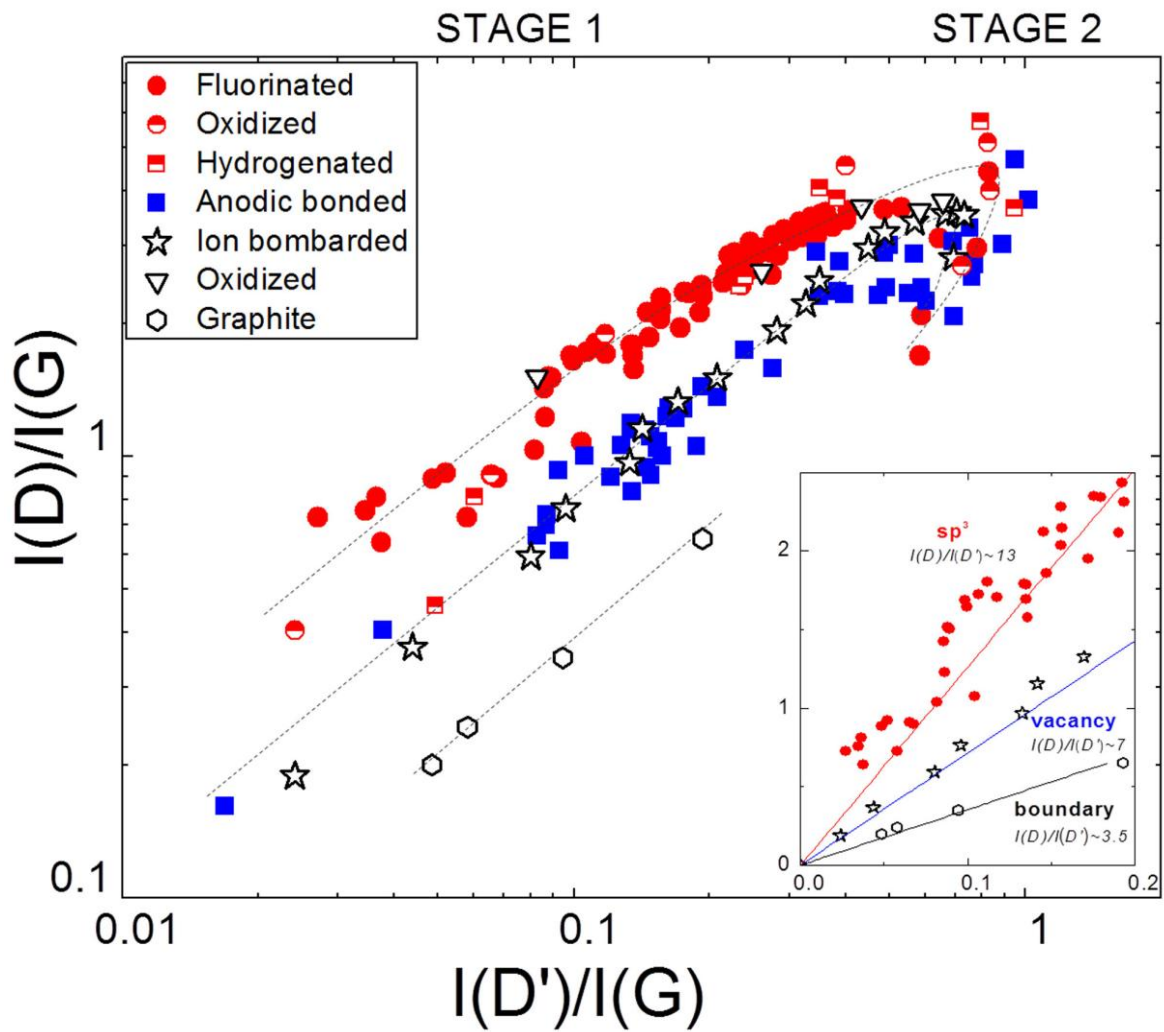

Figure 3 (color on-line) I(D)/I(G) versus ratio I(D')/I(G). Data from literature (open symbols) have been included: ion bombarded graphene,[30,31] oxidised graphene[40] and graphite with different grain sizes.[41] The dotted lines are only a guide for the eyes. The inset shows the linear dependence between the two parameters at low defect concentration, giving different intensity ratio I(D)/I(D') for different type of defects.

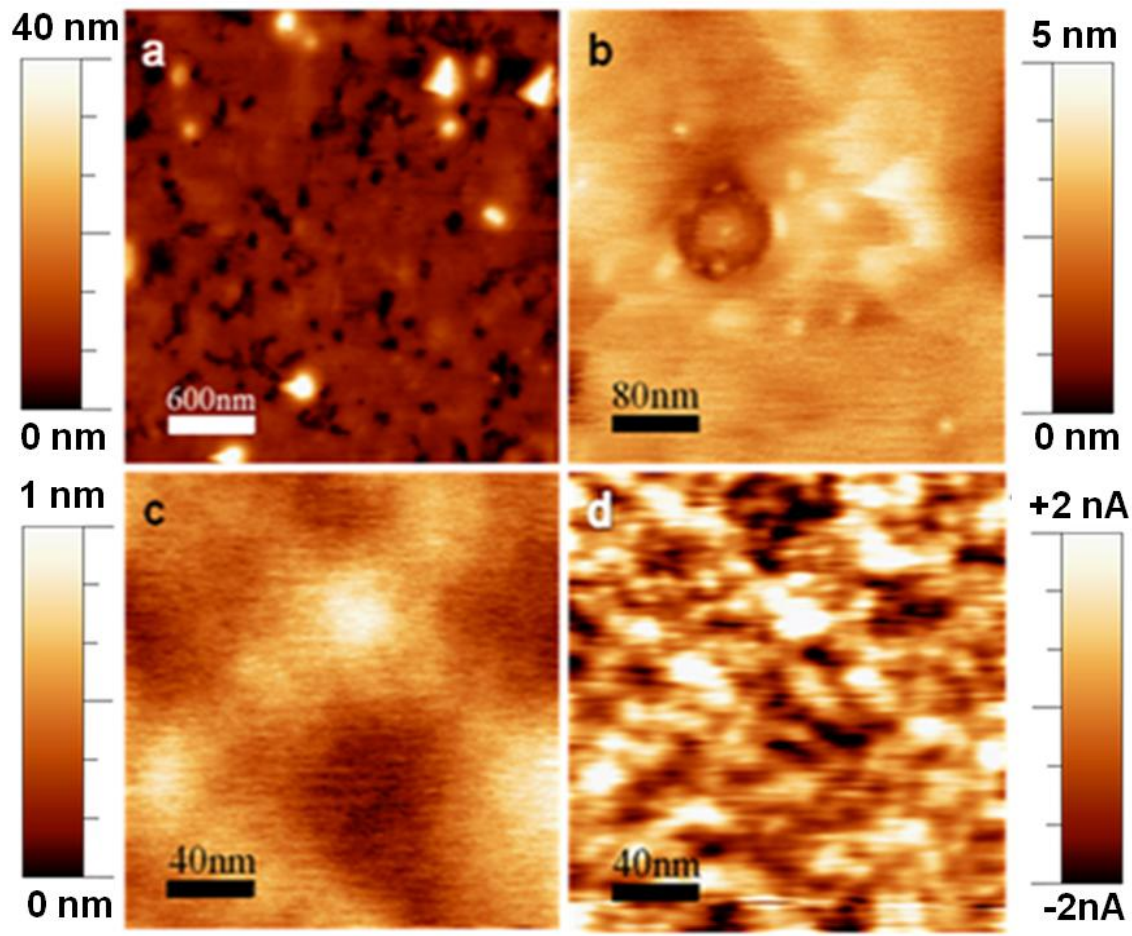

Figure 4 (color on-line) (a) (b) Topography AFM images of pristine defected graphene produced by anodic bonding. (c) Topography and (d) Conductive AFM image on the same area of a fluorinated flake. The current scale shows the variation of the current compared to an average current value obtained from the whole scan. The dark spots correspond to low conductivity areas while the bright spots showed high conductivity.